\documentclass[12pt,preprint]{aastex}

\received{}
\accepted{}
\journalid{}{}
\articleid{}{}

\slugcomment{Accepted for publication in ApJ}

\righthead{}

\begin{document}

\title{Photometric Variability in the Ultracool Dwarf BRI~0021$-$0214:\\ 
Possible Evidence for Dust Clouds}

\author{Eduardo L. Mart\'\i n}
\affil{Institute for Astronomy, University of Hawaii at Manoa,
2680 Woodlawn Drive, Honolulu HI 96822}

\author{Mar\'\i a Rosa Zapatero Osorio}
\affil{Division of Geological and Planetary Sciences,
California Institute of Technology, MS 150-21,
Pasadena, CA 91125}

\author{Harry J.\ Lehto}
\affil{Tuorla Observatory, Turku University, V\"ais\"al\"antie 20,
FI-21500 Piikki\"o, Finland}

\bigskip


\begin{abstract}

We report CCD photometric monitoring of the
nonemission ultracool dwarf BRI~0021$-$0214 (M9.5) obtained
during 10 nights in 1995 November and 4 nights in 1996 August, with
CCD cameras at 1 m class telescopes on the observatories of
the Canary Islands. We present differential photometry of
BRI~0021$-$0214, and we report significant variability in the $I$-band
light curve obtained in 1995.
A periodogram analysis finds a strong peak at a period of 0.84~day. 
This modulation appears to be transient because it is present in the 
1995 data but not in the 1996 data. 
We also find a possible period of 0.20~day, which  
appears to be present in both the 1995 and 1996 datasets. However, 
we do not find any 
periodicity close to the  rotation period expected from the spectroscopic 
rotational broadening ($\le$0.14~day). 
BRI~0021$-$0214 is a very inactive object, with extremely low levels of
H$_\alpha$ and X-ray emission.
Thus, it is unlikely that magnetically induced cool spots
can account for the photometric variability.
The photometric variability of BRI~0021$-$0214 could be explained by the presence
of an active meteorology that leads to inhomogeneous clouds on the surface. 
The lack of photometric modulation at the expected rotational period 
suggests that the pattern of surface features may be more complicated 
than previously anticipated.

\end{abstract}


\section{Introduction}

During the last five years, we have witnessed the discovery of hundreds of
ultracool dwarfs,\footnote{We define  ultracool dwarfs as those with
spectral type M7 or later.} located in the solar neighborhood and in young
open clusters and associations, e.g., \citet{Jones01}.
Detailed study of their properties is still in its infancy,
but some general facts seem to be already well established.

Non-gray model atmospheres have shown that dust forms under the
low-temperature and high-density conditions prevalent in M dwarfs with
$T_{\rm eff}<2,800$~K, e.g., \citet{Tsuji96a} and
\citet{Allard97}.  Dust grains such as calcium titanate (CaTiO$_3$)
and corundum (Al$_2$O$_3$)
have been held responsible for the observed depletion of gaseous TiO from the
atmospheres of the coolest M dwarfs. A new spectral class, designated as L,
has been defined to classify dwarfs cooler than M \citep{Kirkpatrick99, Martin99}.
The optical spectra of late L dwarfs do not show any TiO features,
and they are dominated by broad resonance lines of alkali elements (Cs, K,
Na, Rb).
The $J-K$ colors of late M and L dwarfs are extremely red due to their cool
temperatures
and the greenhouse effect caused by dust opacity, which dissociates
the main molecular absorbers in the $K$-band (CO and H$_2$O).
Dusty atmospheric models are more successful than dust-free models in
reproducing the observed colors of late M and L dwarfs in the field
\citep{Leggett98}, in
the Pleiades cluster \citep{Martin00}, and in very low-mass multiple
systems   \citep{White99, Leinert00}. On the other
hand,  the same models cannot
explain the blue $J-K$ colors of even cooler dwarfs ($T_{\rm eff}<1,200$~K),
such as Gl~229~B, without gravitational settling of
dust grains  \citep{Tsuji96b}. The real atmospheres of ultracool dwarfs are
likely to be
very dynamical, due to turbulent motions related to convection and rotation
\citep{Allard01, marley01}.

Ultracool dwarfs are generally fast rotators.
Rotational broadening ($v$\,sin\,$i$) for about 40 late-M and L dwarfs has
been measured using
high-resolution spectra
\citep {Basri95, Martin97,
Tinney98, Basri00, Basri01}.
The majority of the ultracool dwarfs have $v\sin i>5$ km~s$^{-1}$, with a
trend
for faster rotation toward cooler temperatures. The late L dwarfs also tend
to be
very inactive as judged from the flux of H$_\alpha$ emission \citep{Gizis00}.
Chandra observations of the M9 dwarf LP944-20 have set an upper limit to the
quiescent X-ray emission that is two orders of magnitude lower than typical
X-ray emission
levels in M0--M6 dwarfs \citep{Rutledge00}.

BRI~0021$-$0214 is an ultracool dwarf (M9.5) that was discovered during the
course of
a survey for high-redshift quasars because it has an extremely red $B-R$ color
\citep{Irwin91}.  Among the known very late M and L dwarfs,
BRI~0021$-$0214 is one of the fastest rotators with
$v\sin i=37.5\pm2.5$~km~s$^{-1}$ \citep{Basri01}.
Moreover, it is a very inactive dwarf because it usually does not show a
detectable
H$_\alpha$ emission line \citep{Basri95}, although a weak flare has been
observed once \citep{Reid99}. Even at the maximum of the flare, the level
of  activity was 3 times lower than the mean quiescent level in mid-type M
dwarfs.
The flare recurrence rate was estimated to be $<$7\% of the time.
\citet{Neuhauser99} reported a nondetection of X-rays from BRI~0021$-$0214 in
a pointed ROSAT/HRI observation of 106 kiloseconds, implying
 $\log L_x/L_{\rm bol}<-4.68$, which is one of their most sensitive upper
limits
to X-ray emission from brown dwarf candidates.

The fast rotation of BRI~0021$-$0214, originally reported by \citet{Basri95},
prompted us to start a program of CCD monitoring in 1995. Earlier results
of a   program
of CCD observations of M7--M9 field dwarfs have been reported in
\citet{Martin96}. Thermal inhomogeneities in the atmosphere,  such as
magnetic cool spots
or massive cloud systems, could be modulated with the  rotation period,
which is expected
to be shorter than 0.135~day for a typical radius  of 0.1~$R_\odot$. Thus
during one
observing night, the object completes a few   rotation cycles. This paper
reports on the
results of monitoring the behavior  of BRI~0021$-$0214 in the $I$-band over
10 nights in
1995 and 4 nights in 1996.  In \S 2 we describe the data acquisition,
reduction, and
analysis techniques.  In \S 3 we present the light curves that result from the
differential photometry.  We discuss the results and
implications in \S 4, and our conclusions and directions for future
research in \S 5.

\section{Observations and Data Analysis}

CCD images of BRI~0021$-$0214
were obtained on the nights of 1995 November 8--15 and 22--23
with the 1 m Jacobus Kapteyn Telescope (JKT) at the Observatorio del
Roque de los Muchachos (La Palma), and on the nights of 1996 August 19--22
with the 0.82 m telescope (IAC80) at the Observatorio del Teide (Tenerife).
We used standard broadband $I$ filters. The JKT was equipped with
a $1024 \times 1024$ pixel
Tektronix camera mounted on the Cassegrain focus, providing a
field of view of 32.3 arcmin$^2$ (0\farcs333 pix$^{-1}$). The IAC80
was equipped with a $1024 \times 1024$ pixel
Thomson camera mounted on the Cassegrain focus, providing a
field of view of 54.4 arcmin$^2$ (0\farcs433 pix$^{-1}$).
All nights were photometric with
the exception of 1995 November 15  and 1996 August 21--22, which had some
patchy thin
 cirrus. Exposure times ranged from
300~s to 500~s. We adjusted the exposure time
to compensate for extinction due to high airmass or cirrus clouds.
The seeing of the images varied from 1\farcs0 to 2\farcs5, and the
airmasses at which they were taken ranged from 1.18 up to 1.80.
All the CCD frames were bias subtracted and flat fielded using packages
within the IRAF environment.\footnote{IRAF is distributed
by National Optical Astronomy Observatory, which is operated by the Association
of Universities for Research in Astronomy, Inc., under contract with the
National Science Foundation.}

We performed aperture photometry on BRI~0021$-$0214 and on five other
stars in the field  using ``vaphot,'' an IRAF script
written by Hans Deeg \citep{Deeg98}.
We verified that the two brightest stars in the field of view are not variable
on the timescale
of our observations, and we have used them as reference stars.
They are shown in Figure~1. Images were first aligned, and the
nonoverlapping parts were removed. Changes in the seeing conditions
and in the telescope focus throughout a night implied variations in
the size of the point-spread function (PSF) of the stellar sources
from frame to frame. To take this into account we derived an average
PSF of the three stars by fitting a circular Gaussian on each frame.
All  three sources comply with the requisites for a good definition
of the PSF. Additionally, to obtain high signal-to-noise
aperture photometry, we determined optimum apertures for each star
according to \citet{Howell89}. All apertures were expressed in multiples
of the FWHM of the average PSF, and the sky intensity was defined as
an outer ring 1.5 pixels wide. These multiples were kept constant
from night to night. Finally, instrumental magnitudes were computed,
and relative magnitudes (i.e., differential photometry) of
BRI~0021$-$0214 with respect to the sum of the flux of the two
reference stars were derived. The 1\,$\sigma$ accuracy of the
differential photometry is $\pm$0.006\,mag and $\pm$0.003\,mag for the
IAC80 and JKT data, respectively.

\section{Results}

Figure~2 displays the light curve of BRI~0021$-$0214 obtained at the JKT in
1995 November.
The photometry of reference star 1 with respect to reference star 2 is also
shown.
The target presents a larger photometric dispersion than the two reference
stars.
In Figure~3, we show the standard deviation ($\sigma$) associated with the
total JKT differential light curve of
BRI~0021$-$0214 and five other stars  (including the references 1 and
2) in the field of view of our CCD images.
We call this quantity $\sigma_I$ in analogy to Bailer-Jones \& Mundt
 (2001). The $I$-band photometry was calibrated
using the DENIS $I$-band magnitude of reference star 1 (X. Delfosse, 2000,
private communication).
The value of $\sigma_I$ increases toward fainter magnitude, as expected.
We find a rather tight relationship between
$I$-band magnitude and $\sigma_I$ value for the JKT data ($\sigma_I=0.17144
- 0.028001 \times I + 0.001154 \times I^2$)  with a scatter of 0.0006 mag (1~$\sigma$).
The amplitude of photometric variability in the JKT data of 
BRI~0021$-$0214 is 0.0182 magnitudes. This value is 14~$\sigma$ away from the
relation between I-magnitude versus $\sigma_I$ for the reference stars.

Figure~4 displays the total light curve
of BRI~0021$-$0214 obtained at the IAC80 in 1996 August. The scatter in the
differential photometry of BRI~0021$-$0214 in the
IAC80 data set is $\sigma_I=0.006$ mag, which is a factor of three  
lower than that in the JKT data set ($\sigma_I=0.018$ mag). On the other hand, 
 the photometric errors in the IAC80 data are
larger due to the smaller aperture of the telescope and poorer image quality.
In Figure~5 we plot the $\sigma_I$ versus $I$-band magnitude
(DENIS system). The relation between these two quantities is not as
good as for the JKT data. The standard deviation of the fit shown in Figure~5 is 
0.00316 mag (compare with 0.0006 mag for the fit shown in Figure~3). 
This is probably because there are large PSF
variations across the CCD frames taken with the IAC80.
BRI0021--0214 and reference star 1 have a better PSF than the
average of the other reference stars because they are closer to the
center of the chip. Our differential photometry was obtained by
averaging the fluxes of the reference stars, and this procedure
gives more weight to the brighter sources. The reference star 1 is
the brightest source. Figure~5 indicates that photometric variability of 
BRI~0021$-$0214 with the same amplitude as in the 1995 JKT dataset 
($\sigma_I=0.018$ mag) can be ruled out in the IAC80 dataset with a confidence 
of 2.5~$\sigma$. Thus, the photometric variability of BRI~0021$-$0214 in the I-band 
appears to have decreased from 1995 November to 1996 August. In the next subsection we will show 
that this effect can be accounted for by a source of periodic modulation which 
was present in 1995  but vanished in 1996.

Since  BRI~0021$-$0214 is redder than the reference stars, we checked for
the presence
of correlations between differential magnitude and airmass or seeing, which, 
if present, might indicate color-dependent differential photometric effects due to refraction 
and scattering in the Earth's atmosphere. We did not find any such
correlation, as illustrated in Figures 6 and 7.
We conclude that the variability of BRI~0021$-$0214 is likely to be intrinsic 
to the object, and not due to changes in the Earth's atmosphere.

\subsection{Periodogram analysis}

We performed a period analysis of the data using a CLEAN algorithm
modified by one of us (H. J. Lehto, 2002, in preparation). For an earlier
version of this method, see
\citet{Roberts87}.  The sampling in both the 1995 November and 1996 August
data sets is such that the
window function shows high sidelobes at small multiples of about 1\,d$^{-1}$.
In principle, these could create ambiguity in the determined period.

Because CLEAN (and any other deconvolution method) contains inherently
degenerate solutions, we ran several solutions with different values of
the gain to get an idea of the stability of the solution. The solutions
for our periodograms remained stable.
In 1995 November,  BRI~0021$-$0214
shows a single strong peak with a frequency of  1.1898\,d$^{-1}$,
corresponding to a period of 0.8405 day.
This peak has a
pseudo-power about 2 orders of magnitude higher than the highest peak in the
comparison star's CLEANed periodogram (Fig.\ 8).
The amplitude of this peak is
0.018 mag, which is similar to the value of $\sigma_I$.
In 1996 August there is no sign of
the 0.8405 day peak, but a new small cluster of peaks
has appeared around 4.8842\,d$^{-1}$ (corresponding to a period of 0.205~day).
These peaks clearly have  a lower amplitude than previously,
and may represent a sinusoid-like variation with a modulated amplitude.
There is also a small possibility that the dominant frequency of this
variation is actually 4.0115\,d$^{-1}$ (corresponding to a period of
0.249~day).
The amplitude of the peak at $P=0.205$ day is
0.007 mag, which is similar to the photometric errors.
The periodogram can, however, find a sinusoid of amplitude below the
photometric uncertainty because the noise is spread over many frequencies
in the power spectrum.

Figures~9 and 10 present the JKT light curve of BRI~0021$-$0214 folded with
a period of 0.84 day.
In the first figure, data points taken at different nights are given
different symbols. In the
second one the data points have been binned in groups of five and a
sinusoid fit is overplotted. We note that around phase=0.5 the data 
seems to deviate from the sinusoidal function. 
Each individual error bar corresponds to the standard deviation of a group of five points.
Figure~11 displays the IAC80 light curve of BRI~0021$-$0214 folded with a
period of 0.20 day.
Data points have been binned by 5, as in Figure~10. The amplitudes of the
sinusoidal fits are
0.022 mag for the JKT and 0.006 mag for the IAC80. After subtracting the
0.84 day sinusoidal
fit from the JKT data, we searched for the 0.2 day period. The JKT residual
photometry
is plotted in phase with $P=0.2$ day in Figure~12. Data points have been
binned in groups of
seven to increase the S/N ratio. The shape of the phased light curve
suggests that the
0.2 day period may be present in the JKT data, although the amplitude is
lower than
the 0.84 day modulation. In fact, the amplitude of the 0.2 day sinusoidal
fit to the
JKT residuals is 0.006 mag, which coincides with the amplitude of the fit
to the IAC80 data.

\section{Discussion}

The presence of the 0.2 day photometric modulation in both the IAC80 and
JKT light curves of BRI~0021$-$0214
suggests that this is a long-lived modulation. Could it be the rotation
period?
The absolute magnitude and $T_{\rm eff}$ of
BRI~0021$-$0214 are $M_{\rm K}=10.15$ and 2,500--2,000~K (Leggett et al. 1998),
respectively.
Another observational constraint is that lithium has been depleted from the
atmosphere
\citep{Basri95}. According to the evolutionary models of
\citet{Chabrier00},
the lack of lithium rules out the possibility that the object is younger
than 0.3~Gyr
and less massive than 0.06~$M_\odot$. Given these constraints, the models
predict that
the radius of  BRI~0021$-$0214 should be $R\le 0.11 R_\odot$. For $R=0.11
R_\odot$,
$\sin i=1$, and $V_{\rm eq}=35$ km~s$^{-1}$, we get a period of 0.14~day,
which is inconsistent
with our periodogram analysis. The radius would have to be 0.14~$R_\odot$
to get a rotation
period of $P=0.2$~day, which is inconsistent with the theory.  We consider
several hypotheses
to explain this discrepancy.

First, we consider the possibility of non-rotational line broadening.
Three different studies have derived $v \sin i$ independently by using
high-resolution optical spectra \citep{Basri95, Tinney98, Basri01}.
They all find values of about 40~km~s$^{-1}$, and we have adopted the most
recent one.
The agreement between $v \sin i$ measurements obtained at different epochs
indicates
that the broadening is not due to an unresolved spectroscopic binary.
BRI~0021$-$0214 does not have an unusual rotation among the sample of late
M and L dwarfs that have
been studied at high spectral resolution. It is unlikely that the
broadening is due to any
line formation effect, which should basically be the same for all dwarfs of
similar gravity
and temperature.

Second, we consider the possibility that the radius is about 30\% larger
than predicted by the models.
Recently, the radius of an extrasolar giant planet has been observed for
the first time.
The planet HD209458b has a radius about 40\% larger than that of Jupiter
\citep{Mazeh00}.
The favored interpretation is that the radius has been puffed up by stellar
insolation
\citep{Burrows00}. The radius of BRI~0021$-$0214, however,
cannot be increased in the same way because of the lack of a close
main-sequence star.
The lack of detected X-ray emission by ROSAT \citep{Neuhauser99}
rules out the possibility that there is a compact relativistic
primary, such as a black hole or neutron star. The effects of rotation have
not yet been
included in models of brown dwarfs. One possible effect in a fully
convective object is
a larger radius and a cooler central temperature \citep{MartinClaret96}.
Preliminary calculations by I. Baraffe (2000, private communication)
indicate, however, that the
rotational distorsion in a fast-rotating 0.1~M$_\odot$ star is small ($\le$1\%).

Our previous discussion has ruled out the possibility of problems with
either the $v$\,sin\,$i$ or
the  theoretical radius. Thus, we must address the evidence that 
BRI~0021$-$0214 presents periodic photometric variability with periods longer than 
the rotation period. 

Cool stars are usually understood by analogy to the Sun. Periodic
photometric variability in cool stars 
arises from the presence of dark spots, where the enhanced magnetic field
suppresses
convective energy transport and cools off the plasma to temperatures below
that of the
photosphere. Early- and mid-M-type dwarfs have a high occurrence of
H$_\alpha$ emission,
light variability, and flares (the BY~Dra syndrome), e.g., \cite{Bopp77,
Herbst89}.  The photometric periods of BY Dra stars are consistent with their
expected rotation periods.  The fraction of the bolometric luminosity that
is emitted in
the H$_\alpha$ line  ($R_{\rm H_\alpha}=L_{\rm H_\alpha}/L_{\rm bol}$) is
$\log~R_{\rm
H_\alpha}=-3.9\pm0.2$  for the active, photometrically variable, BY~Dra stars.
Many of these stars have photometric variability with an amplitude of 0.3
to 0.02 mag in the V-band
\citep{Bopp77, Pettersen87,
Alekseev97}.

BRI~0021$-$0214 is rotating even faster than most BY~Dra stars, but it is
not active.
It has $\log R_{\rm H_\alpha}<-6.3$ \citep{Basri95},
which is more than two orders of magnitude lower than the BY~Dra stars.
Magnetic flares cannot
account for the observed light changes because they happen $<$7\% of the
time and
are too weak to produce continuum light at optical wavelengths \citep{Reid99}.
Thus, it seems unlikely that the source of photometric variability
in BRI~0021$-$0214 is a magnetic field because most of the time there is
no evidence for 
chromospheric or coronal cooling.

\cite{Bailer01} recently presented results of
an $I$-band search for variability in 21 late-M and L dwarfs. Half of their
objects present
evidence for variability with  amplitudes of 0.01 to 0.05 mag
on timescales between 0.4 and 100 hr. CLEAN periodograms found strong
periodicities at about
a few hours in some of the objects, but for other objects the variability
did not seem to
be periodic. For one object with observations obtained 1 year apart, no
common period could be found.
\citet{Bailer01} argued that the periodicities were due to long-lived
surface features
modulated  with the rotation periods, and the absence of periodicities for
other stars was due to
surface features  that evolved in brightness and physical size on
timescales of a few hours to a
few days.  The surface inhomogeneities could plausibly be due to
magnetically induced spots or plages, or  dusty clouds.

If magnetic fields produce surface features, we would expect that the
amplitude of the photometric
modulation may scale with the strength of H$_\alpha$ emission. We could
also expect that photometric
modulation becomes smaller for cooler objects, because they tend to be more
inactive
\citep{Gizis00}. \citet{Bailer01} found that the amplitude of
photometric  modulation does not depend on spectral type from M6 through L5.
In Figure~13 we plot photometric variability ($\sigma_I$) versus H$_\alpha$
equivalent width
(EW$_{\rm H_\alpha}$) of BRI~0021$-$0214 together with L dwarfs for which
Bailer-Jones \& Mundt found
evidence for variability.   For these ultracool dwarfs
there is no correlation between EW$_{\rm H_\alpha}$ and the amplitude of
photometric variability, suggesting that these two observables are not causally
connected.

We conjecture several possibilities to explain the pattern of variability in BRI~0021$-$0214: 
(a) There is differential rotation in  BRI~0021$-$0214, however it would have to be extreme to 
explain the very long periodicity of 0.84~day; (b) There are turbulent regions in BRI~0021$-$0214 
reminiscent of Jupiter's bands and zones. Perhaps bright spots are lagging behind the 
mean rotation of the photosphere; however, the period of 0.84~day seems again to be too extreme 
because Jupiter's zonal velocities are moderate \citep{Smith76}. On the other hand,  
this period may be related to a zonal oscillation. Jupiter's Great Red Spot has a zonal 
oscillation with a period of three months \citep{Solberg69}, and it is possible that an 
ultracool dwarf may have shorter zonal oscillation period; 
(c) The fast rotation of BRI~0021$-$0214 (almost three times faster than
Jupiter) may produce instabilities related to strong Coriolis forces. The instabilities may 
lead to wave motions typical of rotating fluids, similar to inertial waves or Rossby waves e.g., 
\citet{Tritton88}. 
It is possible that the 0.2~day periodicity corresponds to a stable pulsation mode, but the 0.84~day periodicity 
cannot be accounted for by stable pulsations;  (d)  Convection reaches the
optically thin layers of ultracool dwarf atmospheres \citep{Hauschildt99}.
Coriolis forces and convective overshooting may push clouds of dust some distance
beyond the theoretical radius of structural models. Some cloud decks could reach out to
significant altitudes with respect
to the  surface. They would be dragged by rotation, but they could lag
behind the deeper layers.   
The photosphere of BRI~0021$-$0214 may not rotate as a solid body. The 0.2 and 0.84 day periods could 
arise from bright or dark clouds located high above the photosphere.

\section{Conclusions and Future Directions}

We have monitored the brightness of the
ultracool dwarf BRI~0021$-$0214 for 10 nights in 1995 November and 4 nights in 1996 
August using $I$-band CCD imaging.
BRI~0021$-$0214 has shown significant variability, particularly in 1995.
We find a surprisingly long modulation
in the 1995 data set (0.84~day) that cannot be due to rotation. A CLEAN
periodogram of the
1996 data set gives a periodicity of 0.2~day. None of these two periods 
correspond to the expected rotation period ($\le$0.14~day).

Since BRI~0021$-$0214 is a very inactive object with extremely low levels of
H$_\alpha$ and X-ray emission, we favor the interpretation that the source of
surface thermal inhomogeneities required to explain the photometric
variability are clouds rather than magnetically induced spots.
BRI~0021$-$0214 has weak TiO bands, which is a sign of Ti depletion by condensation onto
grains.
The atmosphere of BRI~0021$-$0214 is too warm for water to condense, but
silicate and iron
clouds are expected \citep{Chabrier00, Ackerman01}.
Theoretical models indicate that they play a key role in controlling
opacity and temperature
structure in the atmosphere. 
 The rapid rotation and cool temperature of
BRI~0021$-$0214 could make its surface appear more analogous to Jupiter
than to the Sun.
\citet{Gelino00} have simulated integrated light observations of Jupiter at
400~nm
($B$-band)  made by the {\it Hubble Space Telescope}.
They find that the apparent magnitude could be modulated with Jupiter's
rotation period.
The amplitude of the sinusoidal light curve would be about 0.04 mag, which
is similar to the
amplitudes found by us and \citet{Bailer01}.
\citet{Gelino00} predict that the amplitude would be larger in the thermal
emission
regime.  Thus, near-infrared monitoring of BRI~0021$-$0214 and other
ultracool dwarfs may
be worthwhile.

\citet{Tinney99} reported a search for variability using a tunable
filter
technique to measure changes in the strength of TiO bands.
\citet{Tinney00} discusses more
techniques,  such as long-slit spectra and narrowband imaging. We suggest
that $K$-band
observations could  be particularly valuable to search for changes in the
dust-induced
greenhouse effect on the CO bands.  Optical spectroscopic monitoring could
test the
hypothesis that the clouds are made of dust  because the strengths of the
TiO bands
should be variable. A report of possible variability  of the \ion{Li}{1}
resonance line in
Kelu~1, an L2 dwarf, was made by \citet{Martin98}.  Since the \ion{Li}{1}
line is
formed against a background of TiO bands \citep{Pavlenko00},  the apparent
variability of the  atomic line could be due to variable TiO opacity and/or
variable
continuum opacity. \citet{Kirkpatrick01} reported variability in the L dwarf 
Gl~584~C in the 835--975~nm region in timescales of days and years.  

We have briefly discussed four possible scenarios to explain the variability observed 
in BRI~0021$-$0214: differential rotation; bands and zones moving with different velocities; 
pulsations; and turbulent clouds located high above the photosphere. 
In any case, the variability of BRI~0021$-$0214 is more complicated than previously anticipated. 
Simultaneous multiwavelength observations may shed light on what phenomena are responsible 
for the surprising photometric behaviour of BRI~0021$-$0214.

\acknowledgements

We thank the following people: 
Isabelle Baraffe for providing a digital version of the
theoretical models used
in this work, Xavier Delfosse for giving us DENIS photometric data, 
Ted Simon and an anonymous referee for providing insightful comments, and  
Louise Good for correcting the English language used in the manuscript. 
This paper is based on observations made with the 1 m Jacobus Kapteyn
telescope operated
on the island of La Palma by the Isaac Newton Group at the Observatorio del
Roque
de los Muchachos of the Instituto de Astrof\'\i sica de Canarias (IAC), and
the 0.82 m telescope
operated on the island of Tenerife by the IAC at the Observatorio del Teide. 
Partial funding for this work has been provided by NASA grant NAG 5-9992.

\clearpage

\begin{figure}
\plotone{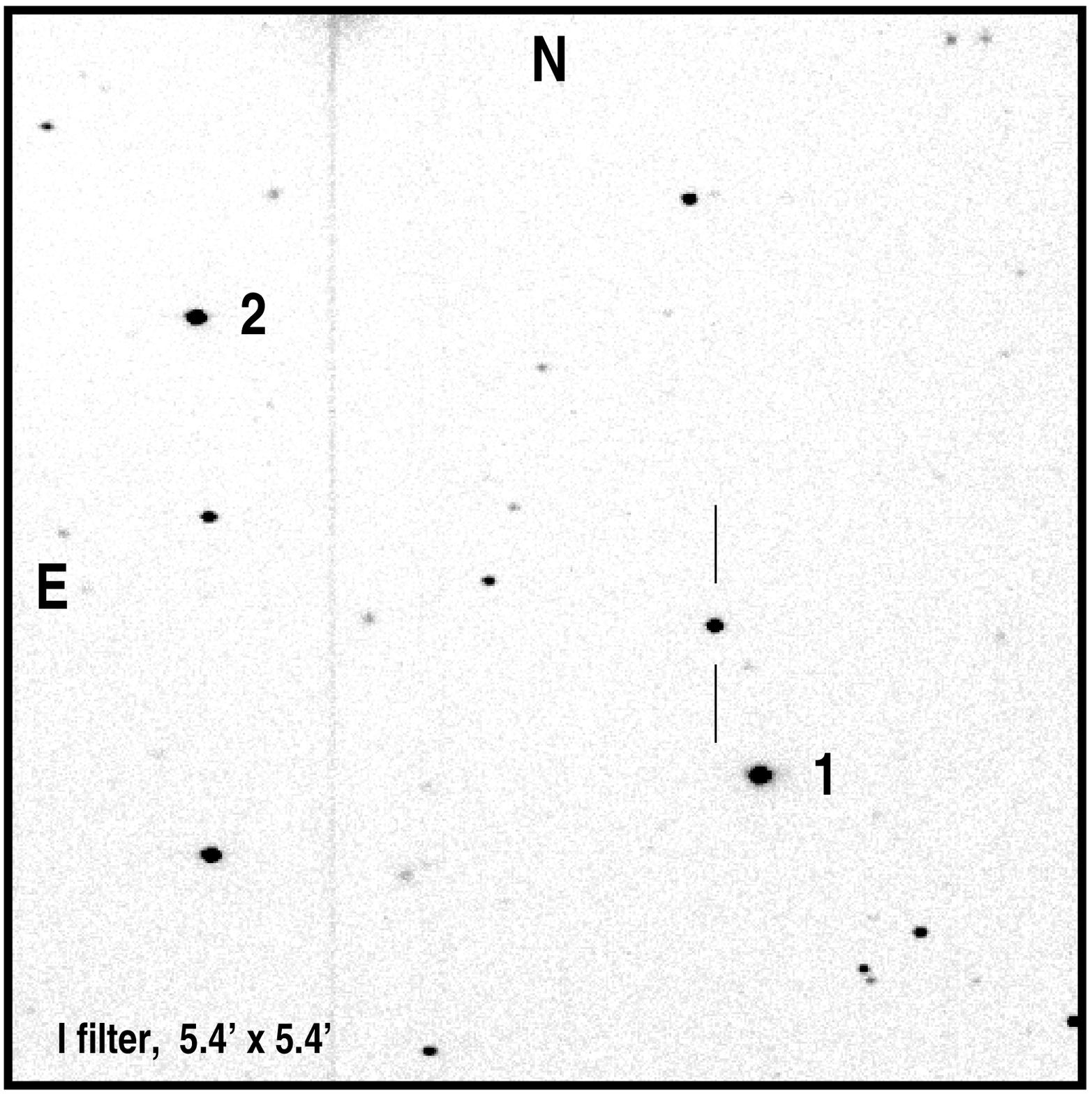}
\caption[]{\label{f1.eps} $I$-band image of BRI~0021$-$0214 obtained with
the IAC80 telescope. The reference stars are indicated.}
\end{figure}

\clearpage
\begin{figure}
\epsscale{0.55}
\plotone{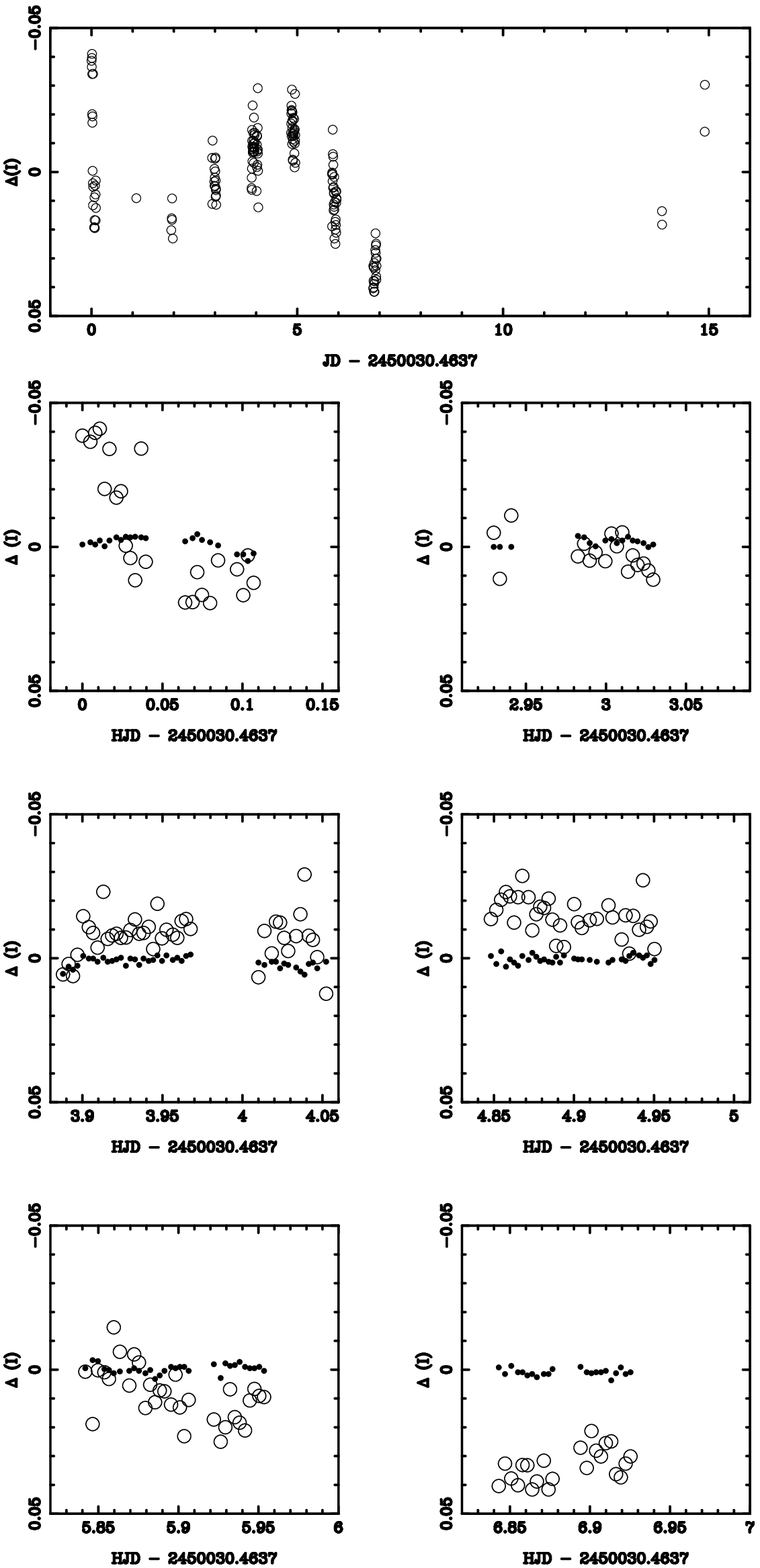}
\caption[f2.eps]{\label{fig2} Complete JKT light curve of
BRI~0021$-$0214. The lower panels display data obtained in individual nights. 
Open symbols are used for BRI~0021$-$0214 and small dots for reference star 1.}
\end{figure}

\clearpage
\begin{figure}
\epsscale{1.0}
\plotone{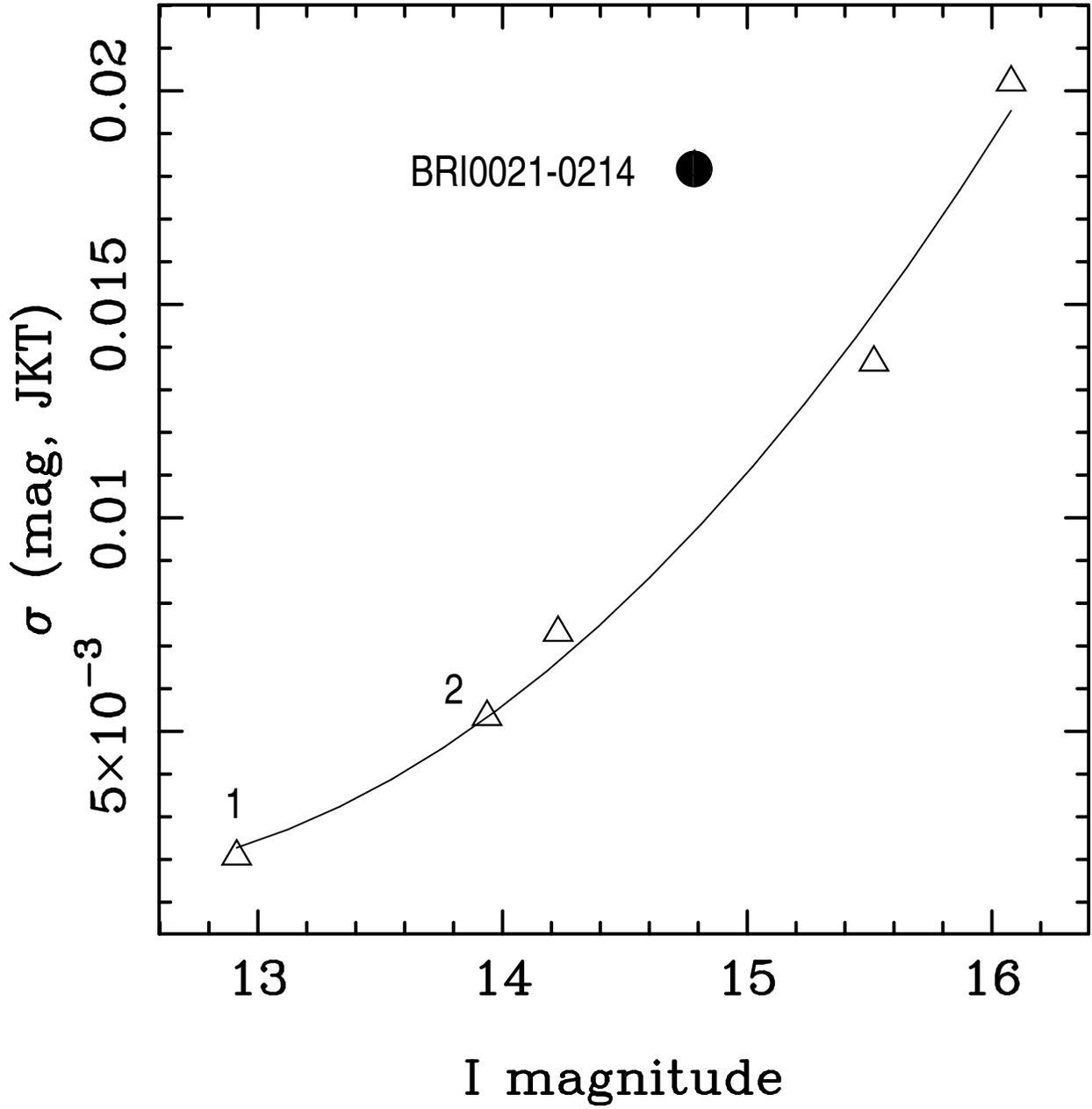}
\caption[f3.eps]{\label{fig3} Standard deviations versus 
$I$-band magnitudes
for BRI~0021$-$0214 (filled circle) and 5 reference stars (open triangles) 
in the JKT database. A second-order polynomial fit to the 
$\sigma_I$ values of the reference stars is shown with a solid line. 
BRI~0021$-$0214 has a $\sigma_I$ value significantly higher than the solid line, 
clearly indicating variability.}
\end{figure}

\clearpage
\begin{figure}
\epsscale{0.65}
\plotone{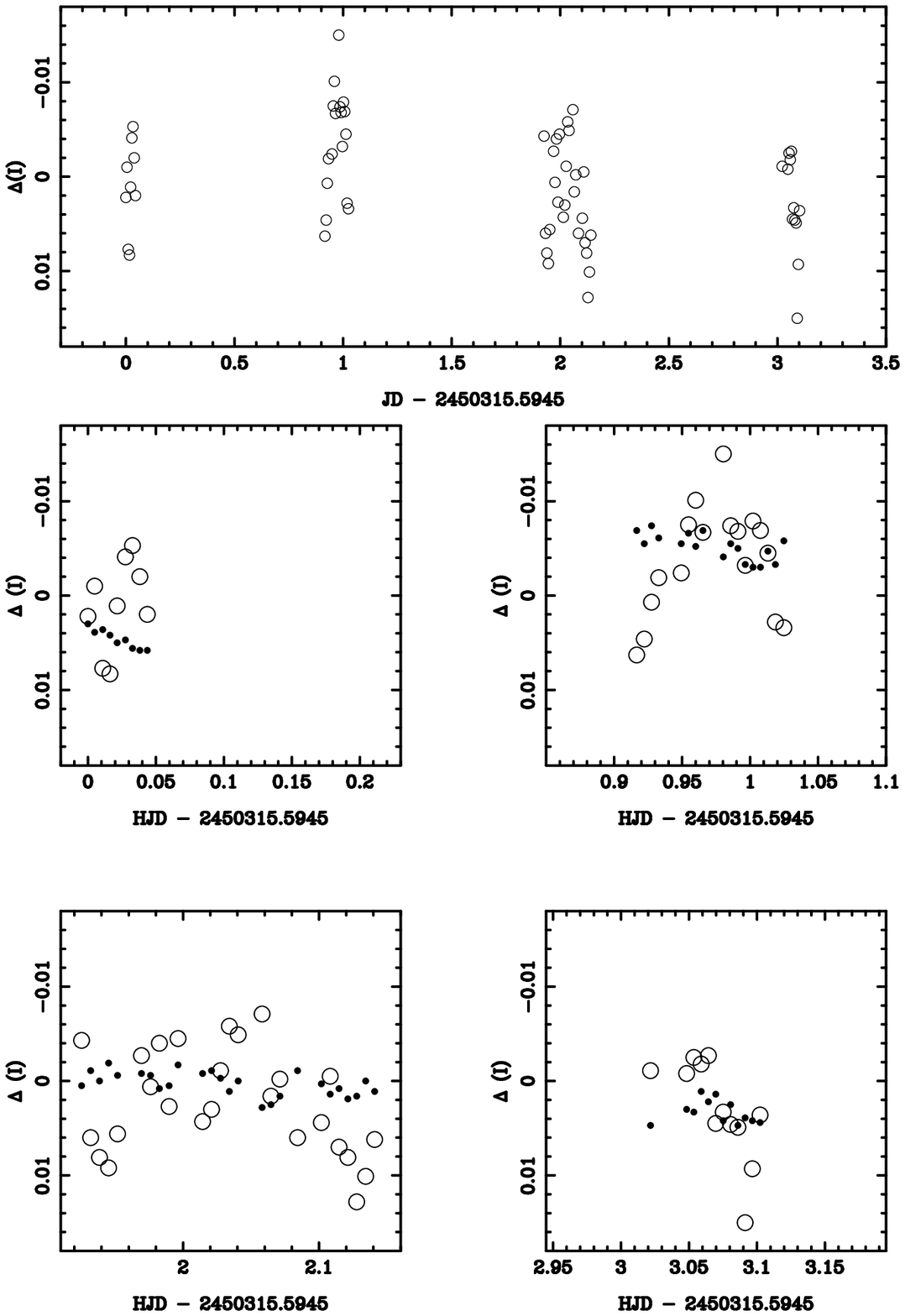}
\caption[f4.eps]{\label{fig4} Complete IAC80 light curve of
BRI~0021$-$0214. The lower panels display data obtained in individual nights. 
Open symbols are used for BRI~0021$-$0214 and small dots for reference star 1.}
\end{figure}

\clearpage
\begin{figure}
\epsscale{1.0}
\plotone{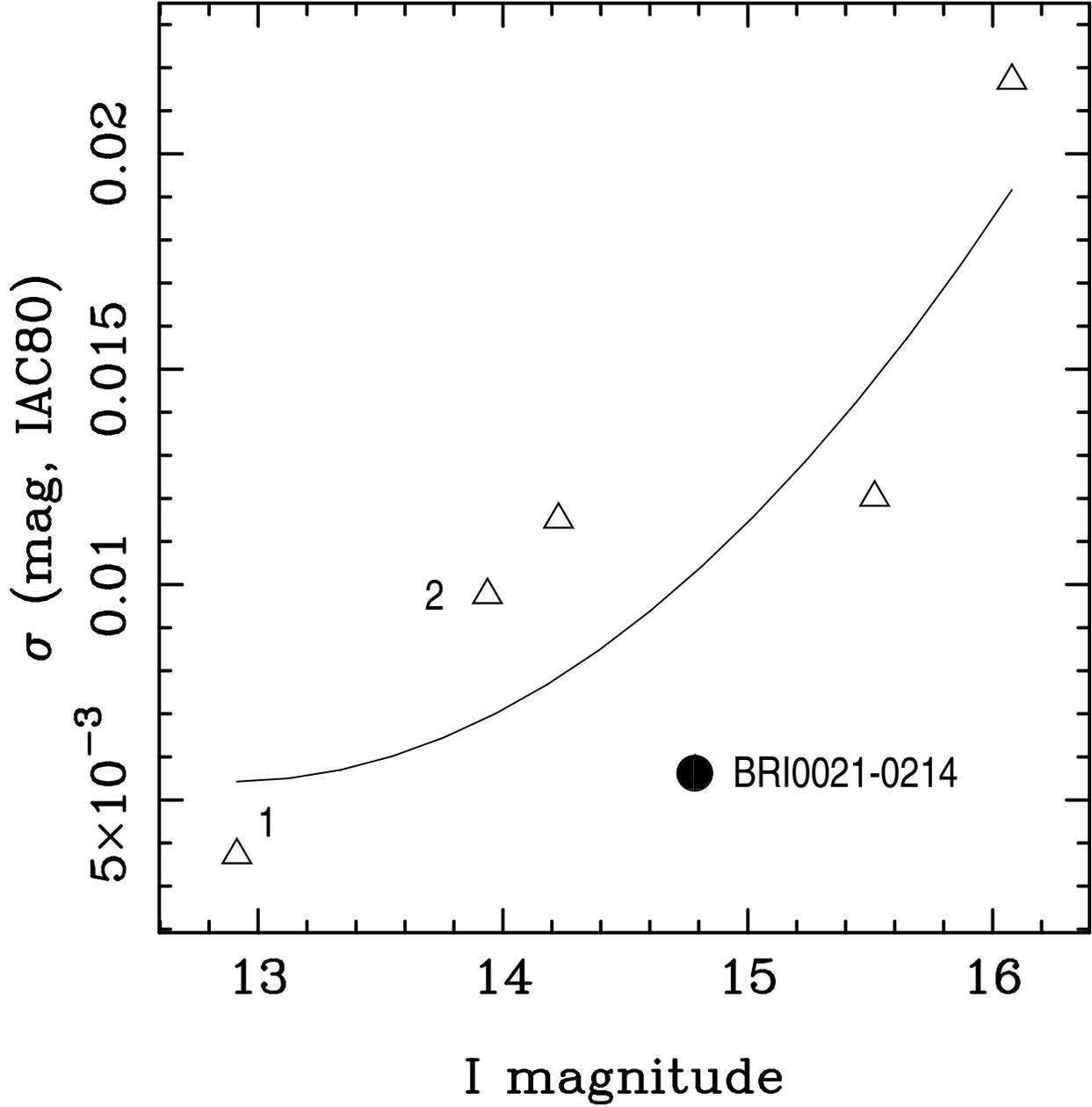}
\caption[f5.eps]{\label{fig5} Standard deviations versus 
$I$-band magnitudes
for BRI~0021$-$0214 (filled circle) and 5 reference stars (open triangles) for 
the IAC80 data. 
A second-order polynomial fit to the 
$\sigma_I$ values of the reference stars is shown with a solid line.}
\end{figure}

\clearpage
\begin{figure}
\epsscale{1.0}
\plotone{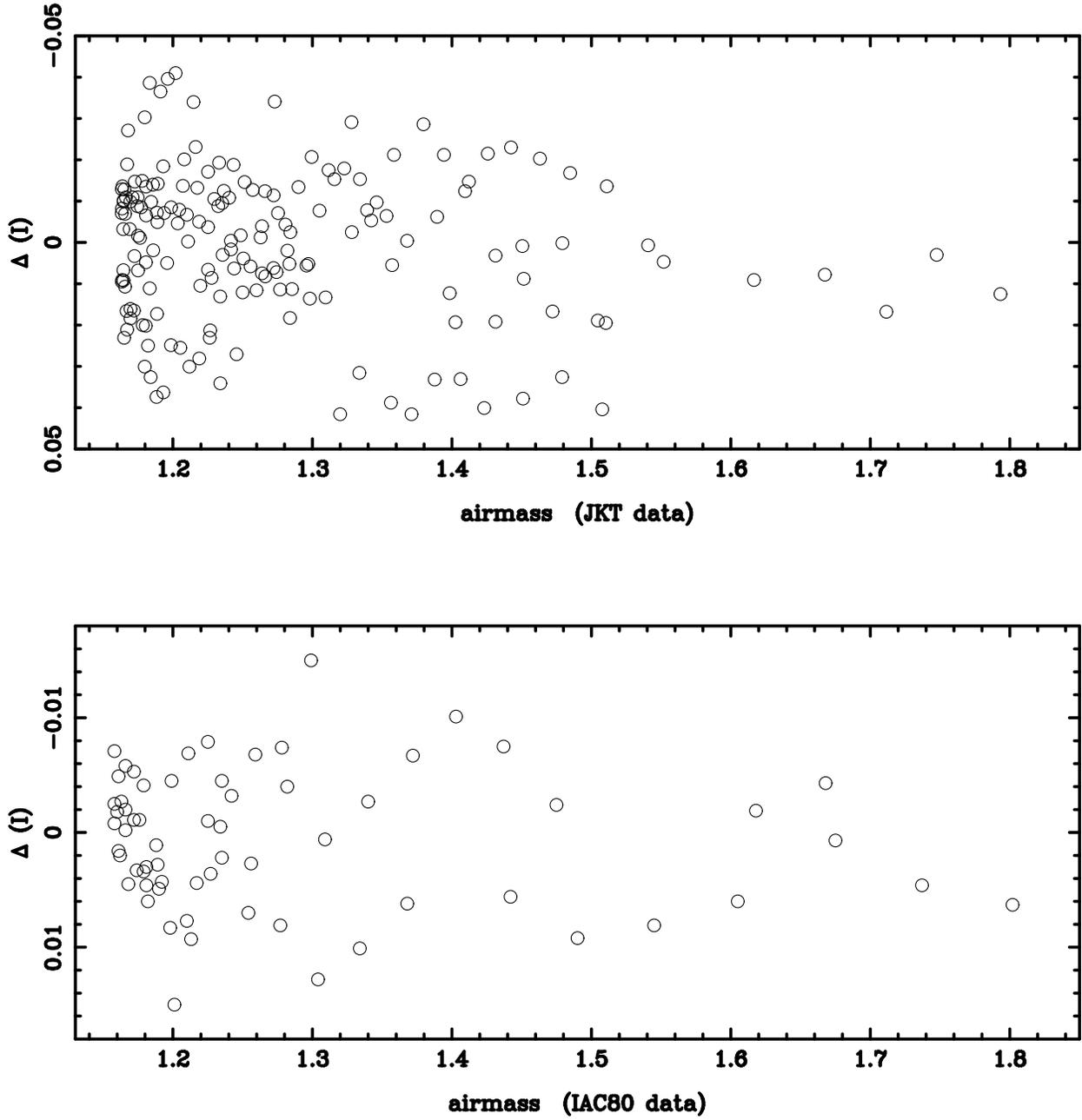}
\caption[f6.eps]{\label{fig6} BRI~0021$-$0214 data as a function of
airmass. No correlation is seen.}
\end{figure}

\clearpage
\begin{figure}
\epsscale{1.0}
\plotone{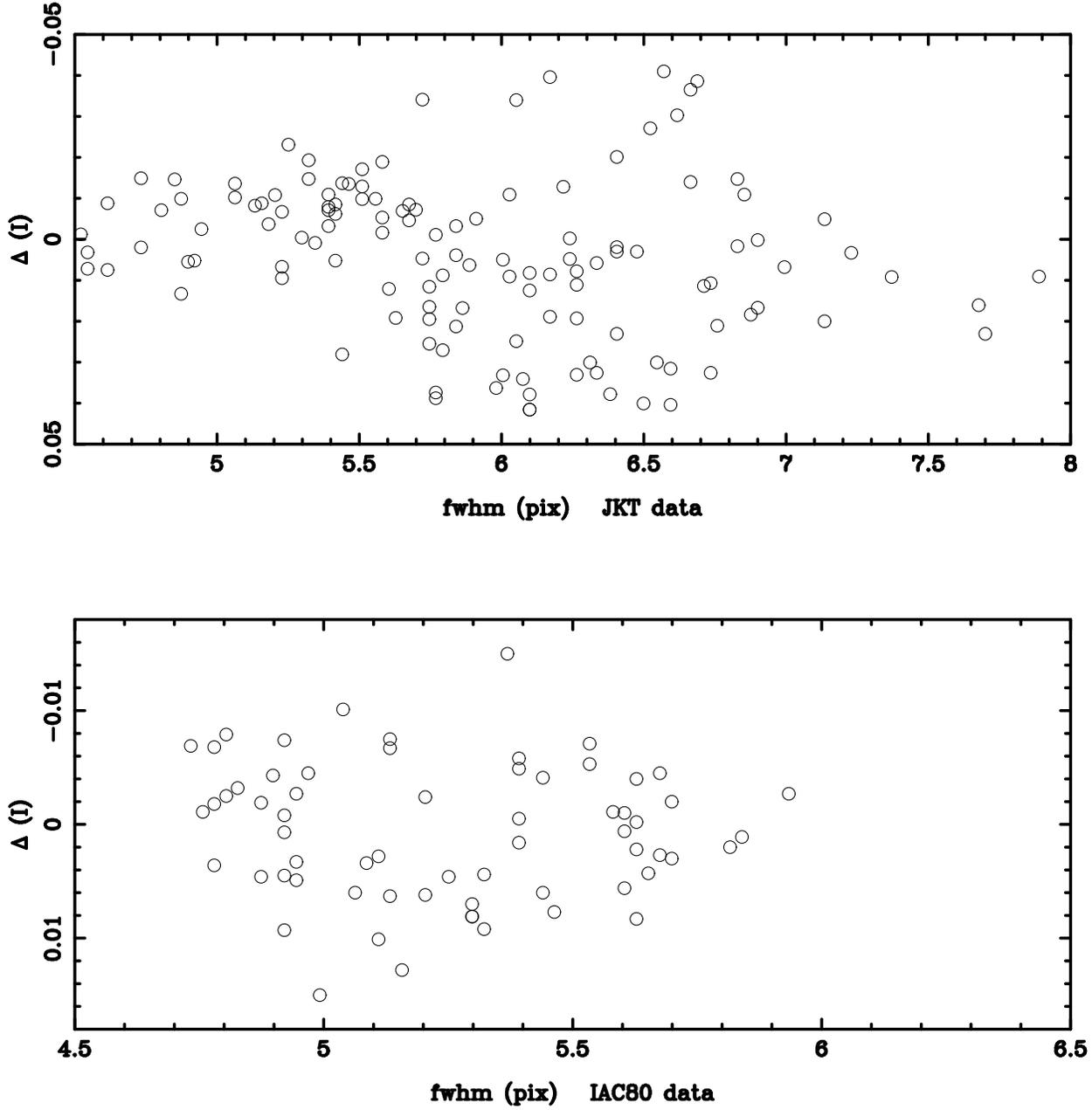}
\caption[f7.eps]{\label{fig7} BRI~0021$-$0214 data as a function of seeing.
No correlation is seen.}
\end{figure}

\clearpage
\begin{figure}
\epsscale{0.8}
\plotone{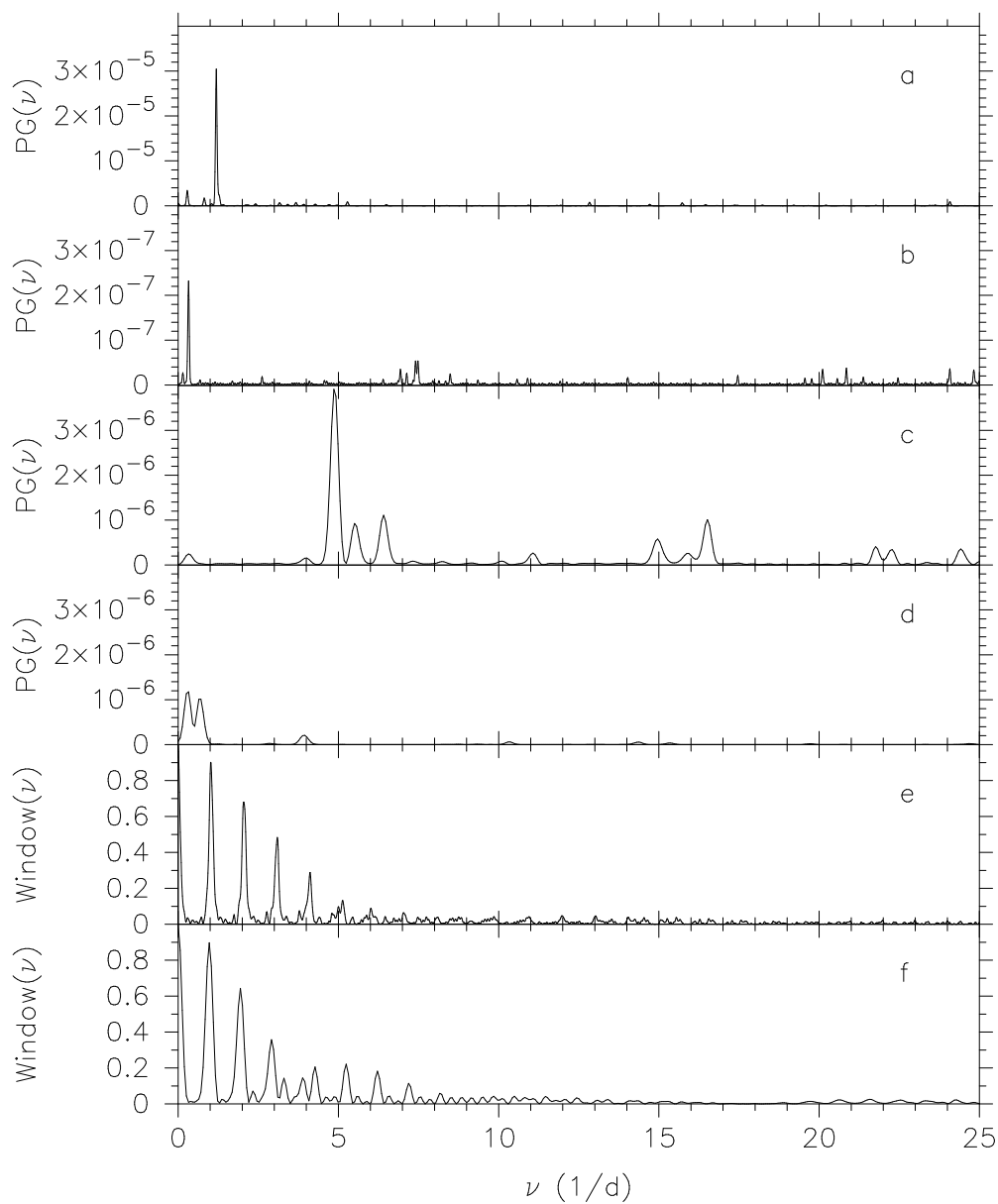}
\caption[f8.eps]{\label{fig8}
CLEANed periodograms of BRI~0021$-$0214 and comparison star 1.
The abscissa in each figure in the inverse of period units of d$^{-1}.$
Panel (a), at the top, is the CLEANed periodogram of BRI~0021$-$0214 during
1995 November (JKT).
Panel (b) is the CLEANed periodogram of comparison star 1 during 1995
November (JKT).
(Note the change in the ordinate representing the pseudo-power.)
Panels (c) and (d) are the respective CLEANed periodograms for BRI~0021$-$0214
and star 1 during 1996 August (IAC80).
Panels (e) and (f) show the window functions of the 1995 November and
1996 August observing sessions.}
\end{figure}

\clearpage
\begin{figure}
\epsscale{1.0}
\plotone{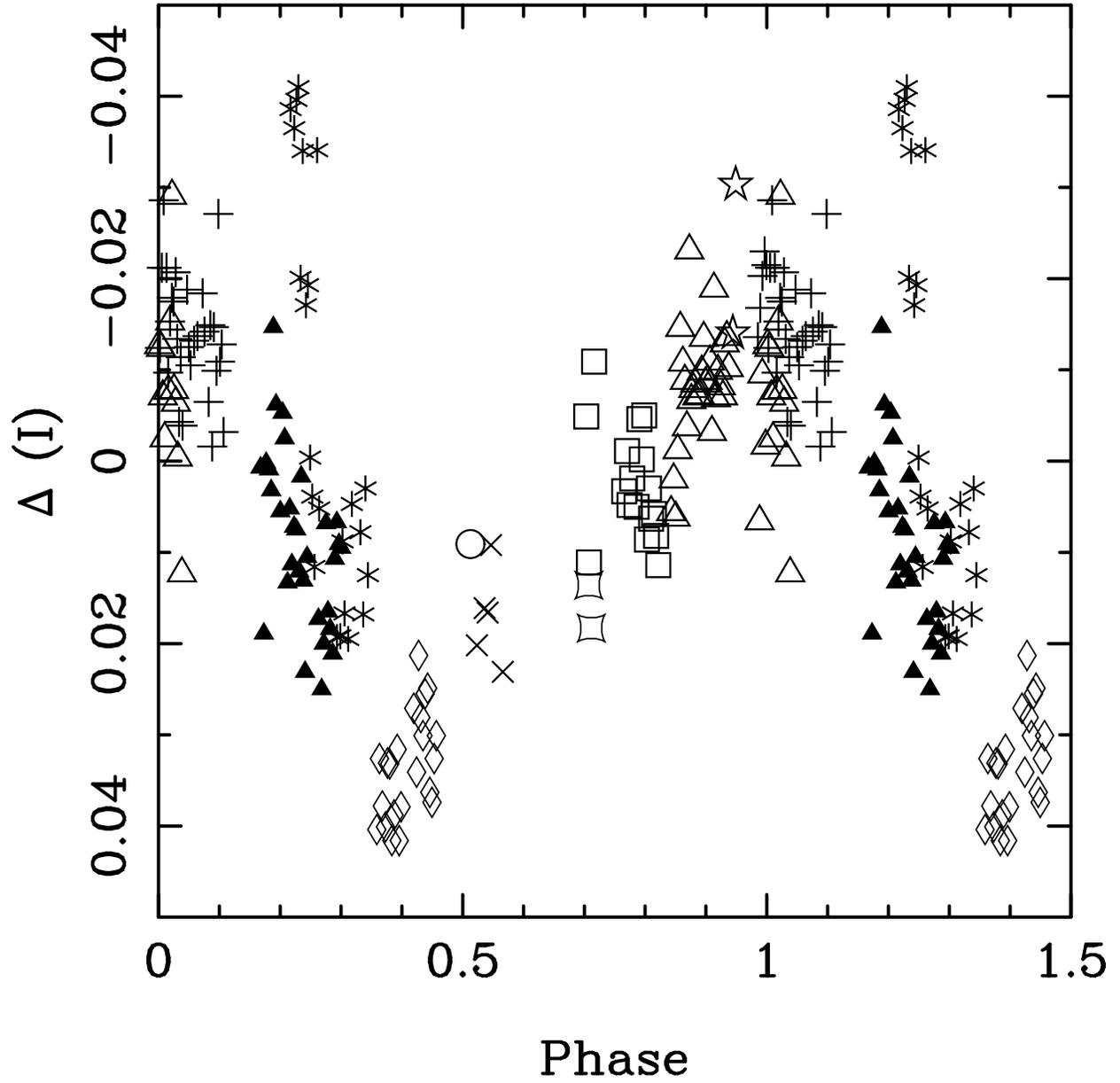}
\caption[f9.eps]{\label{fig9} Phased JKT light curve of BRI~0021$-$0214
with $P=0.84$ day.
Different symbols denote data points obtained in different nights.}
\end{figure}

\clearpage
\begin{figure}
\epsscale{1.0}
\plotone{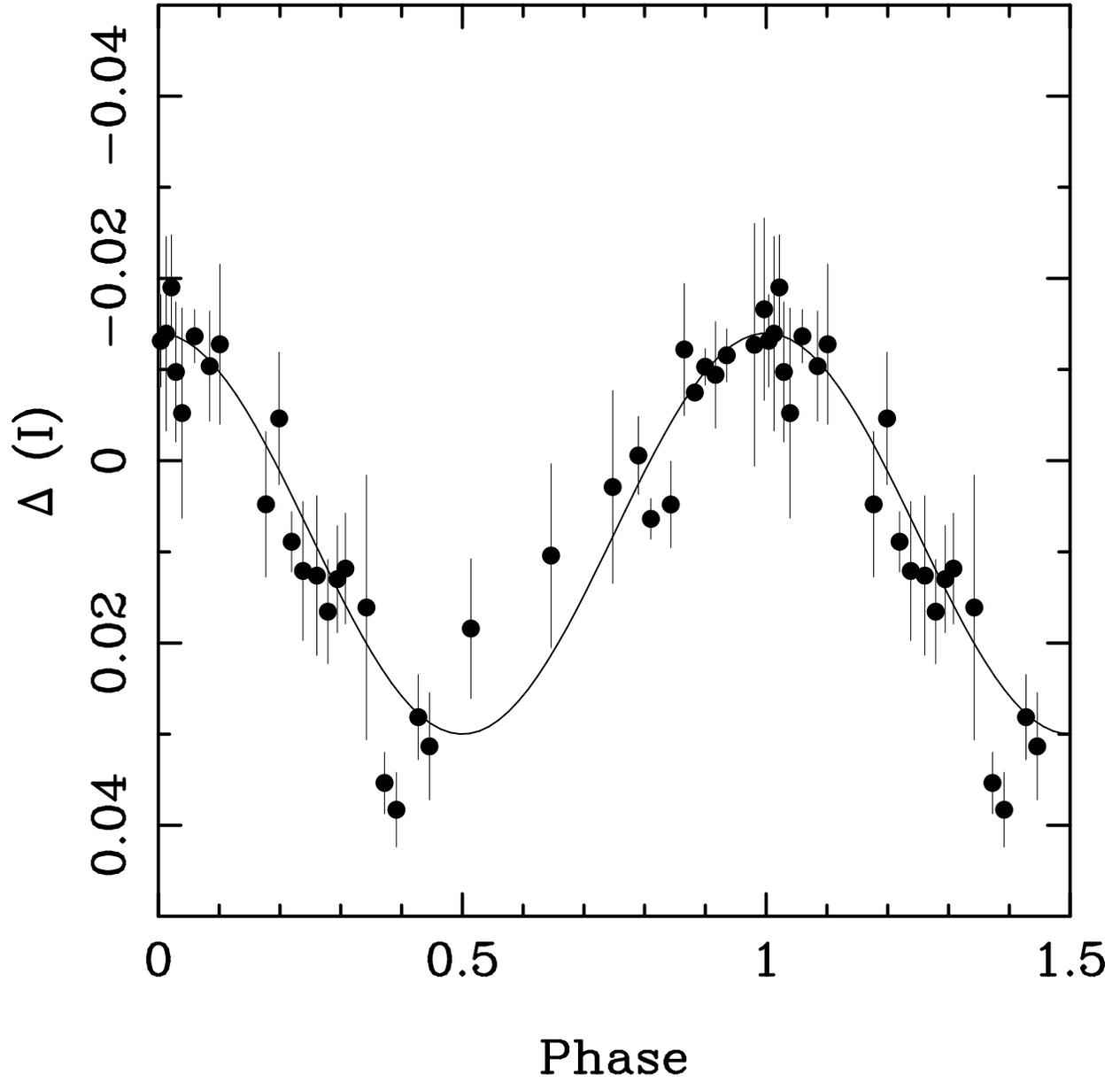}
\caption[f10.eps]{\label{fig10} Phased JKT light curve of
BRI~0021$-$0214 with $P=0.84$ day.
Data points have been binned in groups of five.}
\end{figure}

\clearpage
\begin{figure}
\epsscale{1.0}
\plotone{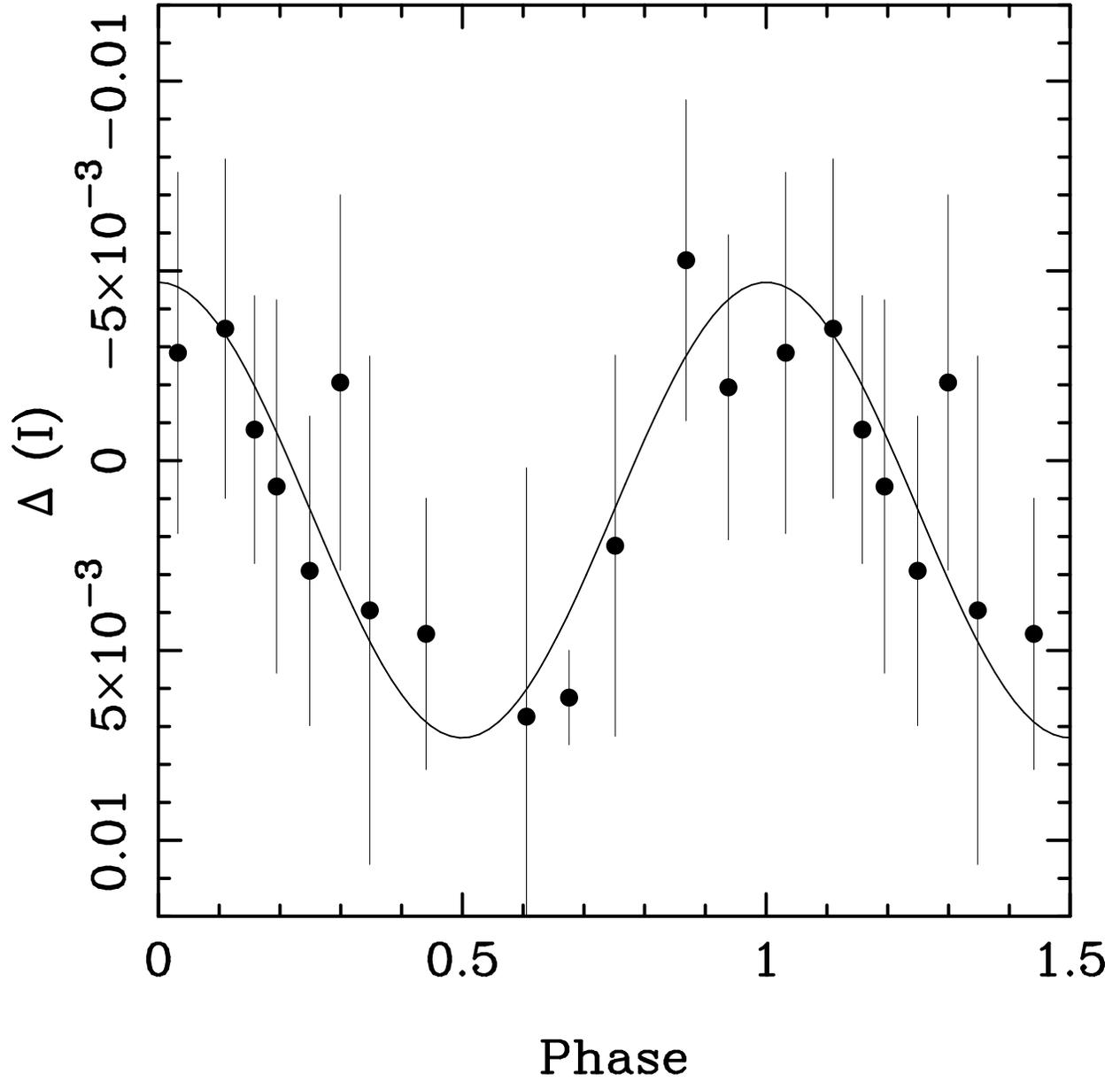}
\caption[f11.eps]{\label{fig11} Phased IAC80 light curve of
BRI~0021$-$0214 with $P=0.20$ day.
Data points have been binned in groups of five.}
\end{figure}

\clearpage
\begin{figure}
\epsscale{1.0}
\plotone{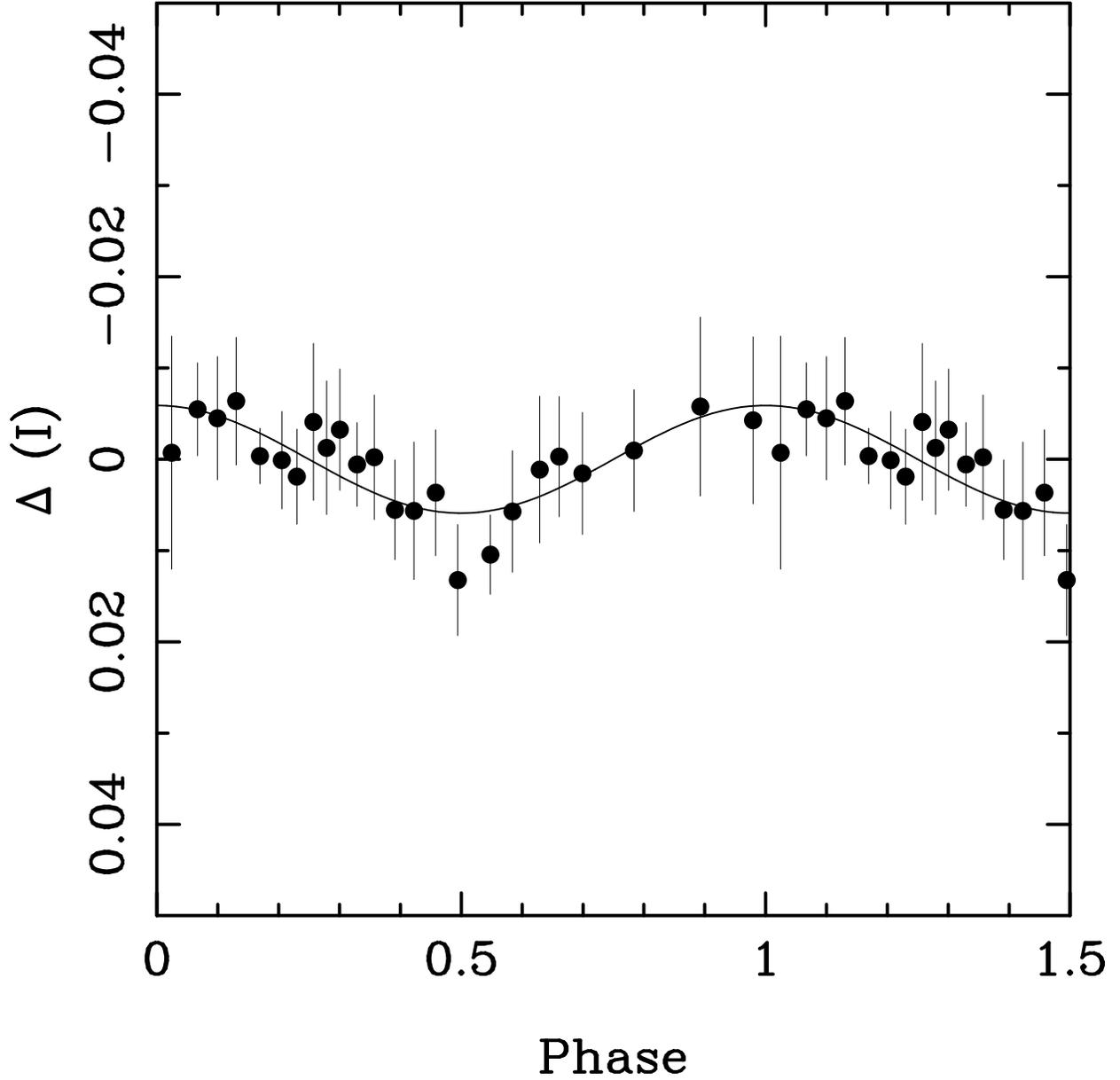}
\caption[f12.eps]{\label{fig12} Phased JKT residual light curve of
BRI~0021$-$0214 with $P=0.20$
day after subtraction of the $P=0.84$ day sinusoidal fit shown in Figure
10.  Data points have been
binned in groups of seven.}
\end{figure}

\clearpage
\begin{figure}
\epsscale{1.0}
\plotone{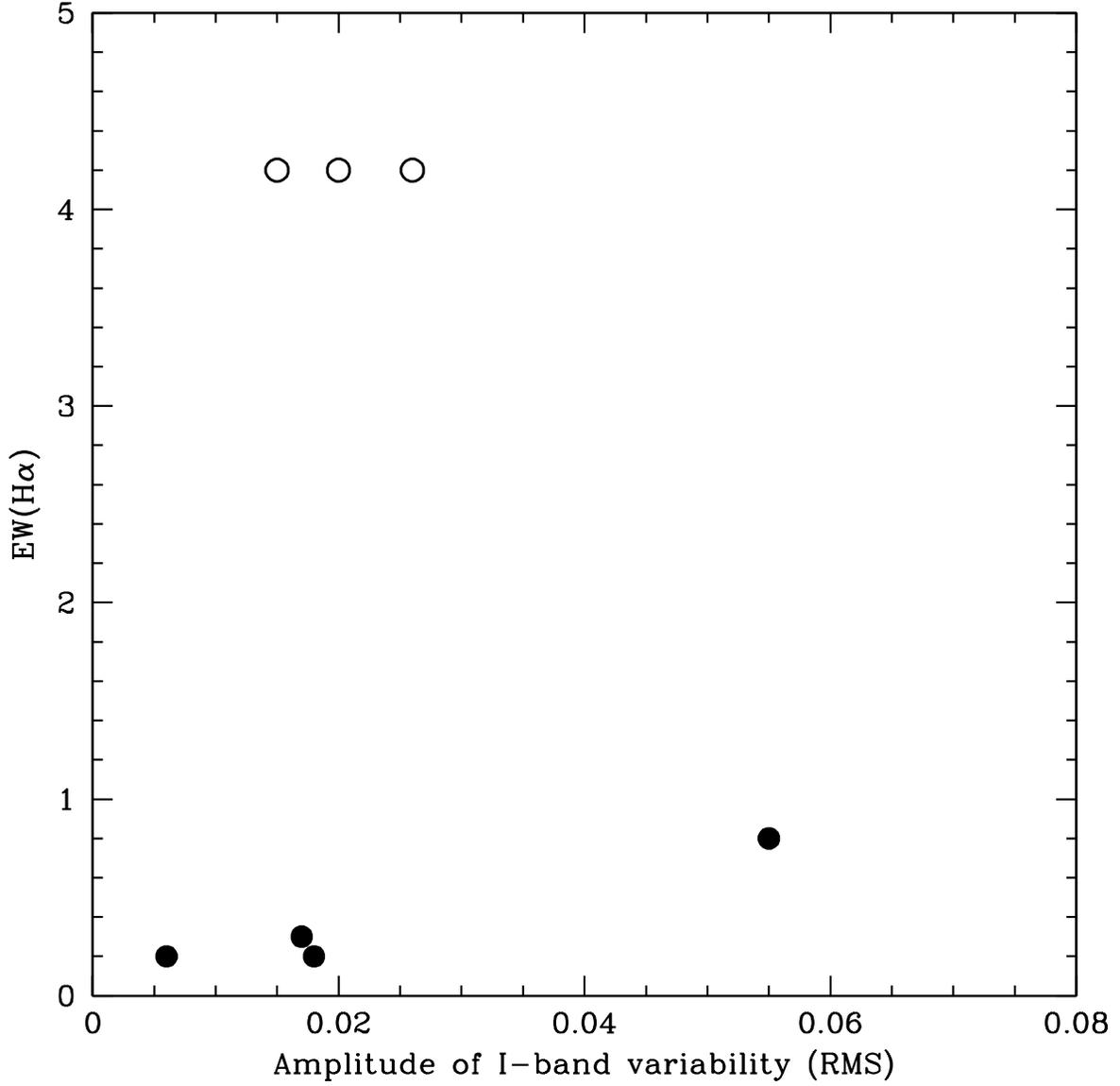}
\caption[f13.eps]{\label{fig13} Amplitude of variability in the $I$-band
of ultracool dwarfs studied in our work and in that of \citet{Bailer01}
versus EW$_{H_\alpha}$. Open circles denote H$_\alpha$ emission detections. Filled
circles
denote H$_\alpha$ emission upper limits.}
\end{figure}

\end{document}